           \documentclass[11pt,titlepage,a4paper]{article}
       \usepackage{amsfonts}

\begin{document}

%New definitions:
%
\def\openone{\leavevmode\hbox{\small1\kern-3.8pt\normalsize1}}%matrix 
% 1
%       sign for the end of proof Q. E. D. in short: qed
%\def\qed{\hbox{\vrule width 6pt height 6pt depth 0pt}} % equivalent
% to
\newcommand{\qed}{\vrule width 6pt height 6pt depth 0pt}
\newtheorem{prop}{Proposition}[section]     % [proposition]
\newtheorem{Lem}{Lemma}[section]            % [ LEMMA ]

\title{\bfseries%
{\huge Normal frames and the validity of the equivalence principle}\\
\vspace{10pt}
{\LARGE I. Cases in a neighborhood and at a point}%
\thanks{To be published in Journal of Physics \textbf{A}:
Mathematical and
general}
}
\author{
Bozhidar Z. Iliev%
\thanks{Department Mathematical Modeling,
Institute for Nuclear Research and Nuclear Energy,
Bulgarian Academy of \mbox{Sciences},
blvd. Tzarigradsko Chauss\'ee 72, 1784 Sofia, Bulgaria}$^,$\thanks
{Internet E-mail: bozho@inrne.acad.bg}
\\\\ \\\\ \\\\ \\\\ \\  \textit{Short title: }
\textbf{\large Normal frames and the equivalence principle. I.}
\\\\ \\
\textit{Classification numbers:} 02.40.Hw, 02.40.Ky, 04.20.Cv,
04.50.+h
}
\date{ }

\maketitle

\begin{abstract}
\noindent
A treatment in a neighborhood and at a point of  the  equivalence
principle on the basis of derivations of the tensor  algebra
over a manifold is given. Necessary and sufficient conditions  are
given for the existence of local bases, called normal frames, in which
the components of derivations vanish in a neighborhood or at a point.
 These
frames (bases), if any, are  explicitly  described  and the problem
of their
holonomicity is considered. In particular, the obtained results
concern
symmetric as well as nonsymmetric linear connections.
\end{abstract}

\section {INTRODUCTION}
\label{I}

\par
Usually in a  local  frame  (basis)  the  gravitational  field
strength is identified with the components of a linear  connection
which may be with or without torsion (e.g. the Riemannian one
in general relativity \cite{1} or the one in Riemann-Cartan space-
times
\cite{2}). This linear connection must be compatible with  the
equivalence
 principle in a sense that there must exist ``local'' inertial,
called also Lorentz, frames of reference (bases) in which the gravity
field strength is ``locally'' transformed to  zero.  Mathematically
this means the existence of special ``local'' basis (or bases)
in which the connection's components vanish ``locally''.  Above  the
words ``local'' and ``locally'' are in quotes as they are not
well defined here, a usual fact for  the  physical  literature
\cite{1}, where they often mean ``infinitesimal surrounding of  a
fixed
point of space-time'' \cite{2}. The strict meaning of ``locally''
may  be
at a point, in a neighborhood, along a path  (curve)  or  on  some
other subma\-nifold of the space-time. The present work deals with
the first
two of these meanings of ``locally'' in which cases
the equivalence principle is considered.

\par
The existence of (local) bases or coordinates in which
the components of linear connections \cite{3,4} vanish at a point
\cite{2,4,5,6,7,8}, along a curve~\cite{5,8} or in a neighborhood
\cite{5,7,8}
have been considered. But with very rare exceptions (see e.g.
\cite{2})
in  the literature  only  the torsion free case has been investigated.
The present work, which is a revised version of~\cite{B-1},
generalizes
these problems to the case of arbitrary derivations  of  the tensor
algebra
over a differentiable manifold (see \cite{4} or Sec. \ref{II}) whose
curvature and torsion are not a priori restricted.

\par
Physically the goal of the paper is to show that  gravity
theories, based, first of all, on linear connections,  are  compatible
with the equivalence principle because of  their  underlying
mathematical structure.

\par
Mathematically the main purpose of this work is to find necessary and
sufficient conditions to be found for the existence of local bases
(coordinates) in which the  components of a derivation (of the tensor
algebra
over a manifold) vanish. If such special bases (frames, called
normal) exist,
the problem of their holonomicity~\cite{5} is considered.

\par
In Sec. \ref{II}  some  notations
and definitions are introduced. Sec.~\ref{III}  deals  with  the
above
problems in a neighborhood and  Sec.~\ref{IV} investigates  them
at a single point. Their connection with the  equivalence principle
is  shown in Sec.~\ref{V}.  Sec.~\ref{VI} contains some concluding
remarks.

\section {DERIVATIONS, THEIR COMPONENTS,
\protect\newline CURVATURE AND TORSION}
\label{II}

\par
Let $D$ be a derivation of the tensor algebra over a manifold
$M$ \cite{3,4}. By proposition 3.3 from ch. I of \cite{4} there exists
unique vector field $X$ and unique tensor field $S$ of type $(1,1)$
such
that $D=L_{X}+S$. Here $L_{X}$ is the Lie derivative along $X$
\cite{3,4}
and $S$ is considered as a derivation of the tensor algebra over $M$
\cite{4}.

\par
If $S$ is a map from the set  of $C^{1}$  vector  fields  into  the
tensor fields of type (1,1) and $S:X\mapsto S_{X}$,  then  the
equation
%
% \begin{equation}
$ D^{S}_{X}=L_{X}+S_{X} $
% \label{1}
% \end{equation}
%
\noindent
defines a derivation of the tensor algebra over $M$ for any $C^{1}$
vector
field $X$ \cite{4}.
Such a derivation will be called an $S$-derivation along $X$ and
denoted for brevity simply by $D_{X}$. An $S$-derivation is  a  map
$D$  such
that $D:X\mapsto D_{X}$, where $D_{X}$ is an $S$-derivation along X.

\par
Evidently (see the above cited theorem from~\cite{4}), every
derivation
of the tensor algebra is an $S$-derivation along  some  fixed
vector field and vice versa.

\par
In this work we shall not be interested on the  concrete  dependence
of $S_{X}$ in $D_{X}=L_{X}+S_{X}$ on X. In it as an  example
of $S$-derivation  only  the  covariant  differentiation will be
considered.
It turns out to be an $S$-derivation linear over functions with
respect to
the vectors along which it acts.

\par
Let $\{E_{i}, i=1,\ldots, n:=\dim (M)\}$ be a (coordinate or
not~\cite{5,6})
local basis (frame) of vector fields in the tangent to $M$ bundle. It
is
holonomic (anholonomic) if the vectors $E_{1}, \ldots , E_{n}$
commute  (do  not commute) ~\cite{5,6}.
%
%%%% The next text (beginning with %Q) is revised (second referee's
report!).
%
%Q Let $T$ be a $C^{1}$ tensor field of type $(p,q)$, $p$ and $q$
%Q being integers or zero(s), with local components
%Q $T^{i_{1}...i_{p}}_{j_{1}...j_{q}}$
%Q with respect to the tensor basis associated with $\{E_{i}\}$.
%Q Here  and below all Latin indices, may be with some super- or
subscripts,
%Q run  from  $1$ to $n:=\dim (M)$. Using the explicit action
%Q of $L_{X}$  and $S_{X}$ on  tensor fields \cite{4} and the usual
summation
%Q rule on repeated  on  different
%Q levels indices, we find the components of $D_{X}T$ to be
%Q %
%Q % \begin{eqnarray}
%Q % & &
%Q $
%Q (D_{X}T)^{i_{1}...i_{p}}_{j_{1}...j_{q}}
%Q =X\left( T^{i_{1}...i_{p}}_{j_{1}...j_{q}} \right) +
%Q \sum^{p}_{a=1}(W_{X})^{i_{a}}_{..k}
%Q T^{i_{1}...i_{a-1}ki_{a+1}...i_{p}}_{j_{1}...j_{q}} +  % \nonumber
\\
%Q % & & +
%Q \sum^{q}_{b=1}(W_{X})^{k}_{.j_{b}}T^{i_{1}...i_{p}}_
%Q {j_{1}...j_{b-1}kj_{b+1}...j_{q}}.           % \label{2} \\
\nonumber
%Q $
%Q % \end{eqnarray} % \noindent
%Q %
%Q  Here $X(f)$ denotes the action of
%Q $X=X^{k}E_{k}$ on  the $C^{1}$ scalar function $f$, i.e.
%Q $X(f):=X^{k}E_{k}(f)$, and if $C^{i}_{kj}$ define the commutators
%Q of the basic
%Q vector fields by $[E_{j},E_{k}]=C^{i}_{jk}E_{i}$ then
%Q $(W_{X})^{i}_{j}:=(S_{X})^{i}_{j} - E_{j}(X^{i}) +
%Q C^{i}_{kj}X^{k}$.
%Q We call $(W_{X})^{i}_{j}$ {\it components} of $D_{X}$.
%Q Evidently
%Q \begin{equation} D_{X}(E_{j})=(W_{X})^{i}_{j}E_{i}. \label{4}
%Q \end{equation}
%Q \noindent Taken as definitions of $(W_{X})^{i}_{j}$ the last two
%Q equations
%Q are equivalent.
%Q
%
Using the explicit action of $L_{X}$  and $S_{X}$ on  tensor
fields~\cite{4}
one can easily deduce the explicit form of the local components of
$D_X T$
for any $C^1$ tensor field $T$. In particular, we have
\begin{equation}
D_{X}(E_{j})=(W_{X})^{i}_{j}E_{i}. \label{4}
\end{equation}
Here  and below all Latin indices, perhaps with some super- or
subscripts,
run  from  $1$ to $n:=\dim (M)$,
the usual summation rule on indices repeated  on  different levels is
assumed, and
$(W_{X})^{i}_{j}:=(S_{X})^{i}_{j} - E_{j}(X^{i}) + C^{i}_{kj}X^{k}$
where
$X(f)$ denotes the action of $X=X^{k}E_{k}$ on  the $C^{1}$ scalar
function
$f$, i.e.  $X(f):=X^{k}E_{k}(f)$
%   the following two lines have been omitted by mistake
and $C^{i}_{kj}$ define the commutators of the basic
vector fields by $[E_{j},E_{k}]=C^{i}_{jk}E_{i}$.
We call $(W_{X})^{i}_{j}$ the {\it components} of $D_{X}$.

\par
The change $\{ E_{i}\} \mapsto \{ E^{\prime }_{m}:=A^{i}_{m}E_{i}\} $,
$A:=[ A^{i}_{m}]$  being  a
nondegenerate  matrix  function,  implies  the  transformation  of
$(W_{X})^{i}_{j}$ into (see (\ref{4}))
%                           \begin{equation}
$(W^{\prime }_{X})^{m}_{l}=(A^{-1})^{m}_{i}A^{j}_{l}$
$(W_{X})^{i}_{j}+(A^{-1})^{m}_{i}X(A^{i}_{l})$.     %\label{5}
%                           \end{equation}
%                           \noindent
Introducing the  matrices
$W_{X}:=[ (W_{X})^{i}_{j}] $
and
$W^{\prime }_{X}:=[ (W^{\prime }_{X})^{m}_{l}] $
and putting
$X(A):=X^{k}E_{k}(A)=[ X^{k}E_{k}(A^{i}_{m})] $, we get
%
%   \addtocounter {equation} {-1}
%   \def\theequation{\arabic{equation}'}
%
\begin{equation}
W^{\prime }_{X}=A^{-1}\{W_{X}A+X(A)\}. \label{5prime}
\end{equation}

\par
    If $\nabla $ is a linear connection with local components
$\Gamma ^{i}_{jk}$
(see, e.g., \cite{3,4,5}), then
$\nabla _{X}(E_{j})=(\Gamma ^{i}_{jk}X^{k})E_{i}$
[4]. Hence,  we see from (\ref{4}) that
$D_{X}$ is a covariant differentiation along $X$ iff
\begin{equation}
(W_{X})^{i}_{j}=\Gamma ^{i}_{jk}X^{k} \label {6}
\end{equation}
\noindent
for some functions $\Gamma ^{i}_{jk}$. Due to $D^{S}_{X}=L_X+S_X$
a linear  connection $\nabla $ is characterized  by  the  map
$S:X\mapsto S_{X}=\Sigma _{X},$ $\Sigma _{X}(Y):=\nabla _{X}(Y)-
[X,Y]$,
$[X,Y]=L_{X}Y$ being the commutator of the vector fields $X$ and $Y$
\cite{4}.

\par
Let $D$ be an $S$-derivation and $X$, $Y$ and $Z$ be vector fields.
The
{\it curvature operator} $R^{D}$, if $D$ is a $C^{1}$ derivation
(i.e. $(W_{X})^{i}_{j}$ are $C^{1}$ functions),
and the {\it torsion operator} $T^{D}$ of $D$ are, respectively,
%
%       \begin{equation}
$R^{D}(X,Y):=D_{X}D_{Y}-D_{Y}D_{X}-D_{[X,Y]}$ and
$T^{D}(X,Y):=D_{X}Y-D_{Y}X-[X,Y]$.      %\label{7}
%
%       \end{equation}
%
The $S$-derivation $D$ is {\it flat} (={\it curvature free}) or
{\it torsion free} if, respectively, $R^{D}=0$ or $T^{D}=0$ (cf.
\cite{4}).

For a linear connection $\nabla $ due to (\ref{6}), we have:
%
%                           \begin{equation}
$(R^{\nabla }(X,Y))^{i}_{j} =  R^{i}_{jkl}X^{k}Y^{l}$,
$(T^{\nabla }(X,Y))^{i}     =  T^{i}_{kl}X^{k}Y^{l}$,   %   \label{8}
%                           \end{equation}
%
where \cite{4} $ R^{i}_{jkl} $
%   := - E_{l}(\Gamma ^{i}_{jk}) + E_{k}(\Gamma ^{i}_{jl}) -
%   \Gamma ^{m}_{jk}\Gamma ^{i}_{ml} + \Gamma ^{m}_{jl}\Gamma
%Q ^{i}_{mk} -
%   \Gamma ^{i}_{jm}C^{m}_{kl}$
%
and $ T^{i}_{kl} $
%    := - (\Gamma ^{i}_{kl} - \Gamma ^{i}_{lk}) - C^{i}_{kl}$,
are the components, respectively,  of
the curvature and torsion tensors of $\nabla $.
%
%       The next text is skipped.
%
%   Using (\ref{1}) and $L_{X}Y=[X,Y]$, we get
%   $
%   (R^{D}(X,Y))Z = \{ S_{X}S_{Y} - S_{Y}S_{X} + L_{X}(S_{Y}) -
%   L_{Y}(S_{X})\} Z + S_{Y}[X,Z] - S_{X}[Y,Z] - S_{[X,Y]}Z, \quad
%   T^{D}(X,Y)=S_{X}Y-S_{Y}X+[X,Y].
%   $
%   By means of  (\ref{2}),  we  find:
%   $
%   (R^{D}(X,Y))^{i}_{k} = X((W_{Y})^{i}_{k}) - Y((W_{X})^{i}_{k}) +
%   (W_{X})^{i}_{l}(W_{Y})^{l}_{k} - (W_{Y})^{i}_{l}(W_{X})^{l}_{k} -
%   (W_{[X,Y]})^{i}_{k}, \quad
%   (T^{D}(X,Y))^{i} = (W_{X})^{i}_{l}Y^{l} - (W_{Y})^{i}_{l}X^{l} -
%   C^{i}_{kl}X^{k}Y^{l}.
%$
%
\par

\par
Further we shall look for special bases $\{E^{\prime }_{m}\}$ in
which the components $W^\prime _{X}$ of an $S$-derivation $D$
vanish along some  or all vector fields X.
For this purpose we have  to solve (\ref{5prime}) with respect to A
under
certain conditions. If such bases (frames) exist, they will be called
\textit{normal} as in the theory of linear connections they are so
called~\cite{4} (and, some times, geodesic or
Riemannian~\cite{5,Fock}).

\section {NORMAL FRAMES FOR DERIVATIONS \protect\newline
                         IN A NEIGHBORHOOD}
\label{III}

\par
In this section we shall solve the  prob\-lems  of  exis\-tence,
uniqueness and holono\-mi\-city of basis or bases $\{E^{\prime
}_{m}\}$
in  which  the components of a given ($S$-)derivation vanish in some
neighborhood U. Such frames will be called \textit{normal} in $U$.

\begin{prop}
\label{p1}
In $U$ for an $S$-derivation $D$  there  exists  a
basis $\{ E^{\prime }_{m}\} $ such that $W^{\prime }_{X}=0$ for every
$X$
if and only  if  in $U$  the $S$-derivation $D$ is a flat linear
connection,
i.e. iff $D_{X}$  is  a  covariant differentiation along $X$ with
$R^{i}_{\hbox{jkl}}=0.$
\end{prop}
\par
{\it Proof:} Let us fix a basis $\{E_{i}\}$ in U. The  existence  of
$\{E^{\prime }_{m}\}$ with $W^{\prime }_{X}=0$, due
to~(\ref{5prime}), implies
%
%Q is equivalent
%Q to the one of a matrix  A transforming $\{E_{i}\}$ into
%Q $\{E^{\prime }_{m}\}$, so that by (\ref{5prime}), we have
%Q %
%Q % \begin{equation}
%Q $ 0=W^{\prime }_{X}=A^{-1}(W_{X}A+X(A)). $ % \label{9}
%Q % \end{equation}
%Q %
%Q \noindent Hence
%
$W_{X}=-(X(A))A^{-1}$,   i.e.
$(W_{X})^{i}_{j}=-[X^{k}(E_{k}(A^{i}_{m}))](A^{-1})^{m}_{j}$
which by~(\ref{6}),  means  that  $D$ is  a  linear  connection  with
local   components
$\Gamma ^{i}_{jk}=-(E_{k}(A^{i}_{m}))(A^{-1})^{m}_{j}$.
Putting $W_{X}=-(X(A))A^{-1}$ and using $X(A^{-1})=-A^{-1}(X(A))A^{-
1}$,
we  get
$R^{D}=R^{\nabla }=0$.

\par
Conversely, let $D$ be a flat linear connection in U. Let
$\{E^{0}_{i}\}$
be a basis at $x_{0}\in U$. Define the  vector field $E^{\prime
}_{i}$ so
that its value $E^{\prime }_{i}{{\mid }}_{x}$ at $x\in U$ is obtained
from
$E^{0}_{i}$ by the parallel translation (transport), generated by $D$
\cite{3,4}, from $x_{0}$ to $x$. As $D$ is a flat linear connection,
$E^{\prime }_{i}{\mid }_{x}$ does  not  depend on the path of
transport
and the vector  fields $\{E^{\prime }_{i}\}$  are linearly
independent \cite{4,5,6}, i.e. they form a basis on $U$. It is
holonomic iff
$D$ is torsion free on $U$ \cite{5,6}. By definition of a parallel
translation, the  vectors  of  the basis $\{E^{\prime }_{i}\}$ satisfy
$D_{X}E^{\prime }_{i}=0$, which, when combined with (\ref{4}), implies
$W^{\prime }_{X}=0$.~\qed

\par
The main consequence of proposition~\ref{p1}  is  that  the  (flat)
linear connections are the only $S$-derivations  for  which  normal
frames
exist in neighborhoods. These frames,
if any, are holonomic iff the derivation is  torsion  free
\cite{5,6}.
From (\ref{5prime}) one finds that they are connected
by linear transformations with constant
coefficients. By proposition \ref{p1}
a necessary condition for the existence of  the  considered  special
bases for an $S$-derivation $D$ is its flatness, i.e. $R^{D}=0.$

\par
Let us turn now  to  the  above-considered  problems  for
$S$-derivations $D_{X}$ along a {\it fixed} vector field $X$
when $X{\mid }_{x}\neq 0$ for $x\in U$.

\par
If $\{E^{\prime }_{m}=A^{i}_{m}E_{i}\}$ is a basis with
$W^{\prime }_{X}=0$, then by (\ref{5prime})  its  existence is
equivalent
to the  one of  $A:=[ A^{i}_{m}] $   obeying
$W_{X}A+X(A)=0$. As $X$ is fixed, the values of A
at different points are connected through the last equation iff the
points
lie on one and the same integral curve of $X$.
Let $\gamma _{y}:J \to M, \  J$ being  an ${\bf R}$-interval, be
the integral curve for $X$ passing through $y\in M$,  i.e.
$\gamma _{y}(s_{0})=y$ and
$\dot{\gamma }_{y}(s)=X\left|_{\gamma _{y}(s)}\right. $,
$\dot{\gamma }_{y}$ being the  tangent  to $\gamma _{y}$  vector
field,
for $s\in J$ and a fixed $s_{0}\in $J.
Along $\gamma _{y}$ the equation $ W_{X}A+X(A)=0$  reduces  to
$(dA/ds)\left| _{\gamma _{y}(s)}=
  - W_{X}(\gamma _{y}(s))A(\gamma _{y}(s))\right. $
with general solution
$A(\gamma _{y}(s))=Y(s,s_{0}; - W_{X}\circ \gamma _{y})B(\gamma
_{y})$.
Here the nondegenerate matrix $B$  is independent of $s$ and
$Y=Y(s,s_{0};Z),
\ Z$ being a matrix function of $s$,
is the unique solution of the initial-value problem \cite{9}
${dY\over ds}=ZY,\  Y\left| _{s=s_{0}}=\openone \right. $
with ${\openone}$ being the unit matrix. Thus we have proved

\begin{prop}
\label{p2}
Let $D_{X}$ be an $S$-derivation  along a fixed  vector field $X(\neq
0)$.
Then along the integral curves of $X$  there  exist
bases in which the components of $D_{X}$ vanish.
\end{prop}
\par
Hence normal frames exist also at any point at which $X$ is
defined. Any two such bases
are  connected by a linear transformation
with a matrix $A$ such that $X(A)=0$ (see (\ref{4}) and
(\ref{5prime})).

\section {NORMAL FRAMES FOR DERIVATIONS \protect\newline AT A POINT}
\label{IV}

\par
Here problems analogous to the ones  of  the  previous
section will be investigated in the case describing the  behavior
of derivations at a given point.

\par
At first we consider $S$-derivations with  respect  to  a
{\it fixed vector field}, i.e. we shall deal with a {\it fixed}
derivation.

    \begin{Lem}     \label{Lem4.1}
    Let
\[
A(y)=B - \Gamma _{k}B(x^{k}(y) - x^{k}(x_{0})) + B_{kl}(y)(x^{k}(y)
- x^{k}(x_{0}))(x^{l}(y) - x^{l}(x_{0})),
\]
    where  $y,x_0\in M$, $B=\mathrm{const}$ is nondegenerate matrix,
i.e. $\det B \neq 0$, $\Gamma_k$ are independent of $y$  matrices,
and  the
matrices $B_{kl}$ and their first derivatives are bounded when $y\to
x_0$.
Then there exists a neighborhood  $U$ of $x_0$ in which the change of
the
bases
$\{\partial/\partial x^i\}
\to \{ E_{m}^{\prime}=A_{m}^{i}\partial/\partial x^i\}$
is well defined, i.e. bijective, in $U$, that is  $\det A(y)\neq 0$
for
$y\in U$.
    \end{Lem}

{\it Proof.}
    Putting
\(
U':=\{z:\quad 1+c_k\varepsilon^k(z)>0\}
\)
with $\varepsilon^k(y):=x^k(y) - x^k(x_0)$ and
\(
c_k := (\det B)^{-1}
\left(\frac{\partial}{\partial\varepsilon^k(y)}\det
A(y)\right)_{\varepsilon(y) = 0},
\)
 $\varepsilon(y) := \max_{k} |\varepsilon^k(y)|$,
we fined
\(
\det A(y) = (\det B) \det [\openone -\Gamma_k\varepsilon^k(y) +
O((\varepsilon(y))^2)] =
(\det B) [ 1 + c_k\varepsilon^k(y) + O((\varepsilon(y))^2) ].
\)
Hence, using that $\det B \neq 0$, we can choose a neighborhood
$U\subseteq U'$ of $x_0$ such that
$(\det A(y))/(\det B) > 0$ for $y\in U$.
(E.g., as  $f=O(g)$ for
real functions  $f$ and $g$ means the existence of
$\lambda\in\mathbb{R}_+$ such that $|f|\le\lambda|g|$, we can put
 $U:= \{z:\quad z\in U',\
(\sum_{k}|c_k|)\varepsilon(z) + \lambda((\varepsilon(z))^2)\le 1\}$
for some $\lambda\in\mathbb{R}_+$.)
Consequently $A$ defines a
bijective mapping between $\{\partial/\partial x^i\}$ and
$E_{m}^{\prime}$
in any such neighborhood $U$ of  $x_0$.~\qed %

\textbf{Remark}
There are different local coordinates $\{y^i\}$ normal at $x_0$. For
instance,
one class of normal at  $x_0$ coordinates is separated through the
equation
 $x^i(z) = y^i(z) + b_{jk}^{i}(y^j(z)-x^j(x_0))(y^k(z)-x^k(x_0))$
for $[b_{jk}] =-W_X/2$.
More general classes of normal at $x_0$ bases and (holonomic)
coordinates for
fixed $X$ are described in~\cite[propositions 8 and 9]{B-1}.

\begin{prop}
\label{p3}
Let $x_{0}\in M, \  X$ be a vector field with $X{\mid }_{x_{0}}\neq
0,$ and
$D$ be an $S$-derivation. Then there exist normal at  $x_0$ local
coordinates
$\{y^{i}\}$, i.e. such that $(D_{X}\partial _{y^{i}}){\mid
}_{x_{0}}=0.$
\end{prop}

\par {\it Proof.}
%Q        OLD PROOF
%Q
%Q Let $\{z^{i}\}$ be local coordinates  with $x_{0}=(0,\ldots
%Q ,0)$  and $X=\partial _{z^{1}}$. Then for some $(W_{X})^{i}_{j}\in
%Q {\bf R}$
%Q at $x_{0}$ we  have $D_{X}\partial _{z^{i}}=(W_{X})^{j}_{i}\partial
%Q _{z^{j}}$.  We define $\{y^{i}\}$ by
%Q $z^{i}=y^{i}+b^{i}_{jk}y^{i}y^{k}$,
%Q where $b^{i}_{jk}=b^{i}_{kj}\in {\bf R}$. Then we find at $x_{0}$
%Q $D_{X}\partial _{y^{i}}=((W_{X})^{j}_{i}+2b^{j}_{i1})\partial
%Q _{y^{j}}$
%Q because of $X{\mid }_{x_{0}}=\partial _{z^{1}}{\mid }_{(0,...,0)}$.
%Q Now the proposition follows choosing
%Q $b^{j}_{i1}=-(W_{X})^{j}_{i}/2.$~\qed
%Q
%Q  NEW PROOF FOLLOWS
%Q
%
    Let $A(z)$, $z\in M$ be defined as in lemma~\ref{Lem4.1} for
 $B=\openone,\ B_{kl}=0, \mbox{ and }
(\Gamma_k)_{j}^{i}=-2b_{kj}^{i}=-2b_{jk}^{i}\in\mathbb{R}$.
Then by lemma~\ref{Lem4.1} there is a neighborhood $U$ of $x_0$ in
which
 $\{E_{m}^{\prime}=A_{i}^{m}\partial/\partial x^i\}$
with the matrix $A(z)=\openone+\Gamma_k (x^k(z)-x^k(x_0))$
form a field of bases in $U$.
In it, using~(\ref{5prime}), we find
 $W_{X}^{\prime}(x_0)=W_X(x_0) + \Gamma_kX^k\left.\right|_{x_0}$.
As $X\left.\right|_{x_0}\neq 0$ we can choose $\{x^i\}$ such that
 $X=\partial/\partial x^1$. Now, partially fixing $\{\Gamma_k\}$ by
defining
 $\Gamma_1 = [-2b_{j1}^{i}] = W_X$, we get $W_{X}^{\prime}(x_0) = 0$.
Hence $\{E_{m}^{\prime}\}$ is normal at $x_0$.
Besides, it is holonomic at $x_0$ as
 $[E_{k}^{\prime},E_{m}^{\prime}]\left.\right|_{x_0} =
- 2 (b_{km}^{j} -b_{mk}^{j}) \frac{\partial}{\partial
x^j}\left.\right|_{x_0}
\equiv 0$.
So, there exist local coordinates $\{y^i\}$ in a neighborhood $V$ of
$x_0$
such that
 $E_{k}^{\prime}\left.\right|_{x_0} =
\frac{\partial}{\partial x^k}\left.\right|_{x_0}$.
Evidently, in $V\cap U$ the coordinates $\{y^i\}$ are normal at
$x_0$.~\qed

\par
From (\ref{4}) and (\ref{5prime}) we find that a basis
$\{ E^{\prime }_{m}\} $ in which $W^\prime _{X}(x_{0})=0$
is obtained from $\{ \partial /\partial y^{i}\}, \  \{y^{i}\}$ defined
in the last proof, by  a linear transformation with a matrix $A$ such
that
$(X(A)){\mid }_{x}=0$.
%,  i.e. $A(x)=const+O((x^{k}-x^{k}_{0})(x^{l}-x^{l}_{0}))$.
The holonomicity of these normal frames depends on the concrete
choice of $A$.

\par
If $X{\mid }_{x_{0}}=0$, then a basis $\{E^{\prime }_{m}\}$  in  which
$W^\prime _{X}(x_{0})=0$
exists if  $W_{X}(x_{0})=0$ in some basis $\{E_{i}\}$, so then  every
 basis,
including the holonomic ones, will have the needed property.

\par
Let us now turn our attention to $S$-derivations  with  respect
to {\it arbitrary vector fields}.

\par
The $S$-derivation $D$ is {\it linear} at $x_{0}$  if  for
all $X$ and some (and hence any) basis $\{E_{i}\}$ we have (cf.
(\ref{6}))
$W_{X}(x_{0})=\Gamma _{k}X^{k}(x_{0})$, where the
$\Gamma _{k}$ are constant matrices. This means~(\ref{6}) is valid at
$x_{0}$, but may not be true at $x\neq x_{0}$.

\begin{prop}
\label{p4}
An $S$-derivation $D$ is linear at some $x_{0}\in M$  iff there is a
local
basis $\{E^{\prime }_{m}\}$ in which the components  of $D$  along
every vector field vanish at $x_{0}$.
\end{prop}

\par
{\it Proof:} Let $\{x^{i}\}$ be local coordinates in a neighborhood of
$x_{0}$ and $D$ be linear at $x_{0}$ ,  i.e.
$W_{X}(x_{0})=\Gamma _{k}X^{k}(x_{0})$  for  some $\Gamma _{k}$.  We
search for
$\{E^{\prime }_{m}=A^{i}_{m}\partial /\partial x^{i}\}$ in which
$W^\prime _{X}(x_{0})=0$. Due to (\ref{5prime}) this is equivalent to
$\Gamma _{k}A(x_{0})+\partial A/\partial x^{k}{\mid }_{x_{o}}=0$.
Choosing $A(y)$ as in lemma~\ref{Lem4.1}
%Q
%Q $A(y)=B - \Gamma _{k}B(x^{k}(y) - x^{k}(x_{0})) +
%Q B_{kl}(y)(x^{k}(y)
%Q - x^{k}(x_{0}))(x^{l}(y) - x^{l}(x_{0}))$,
%Q where $B=\mathrm{const}$ is nondegenerate,i.e. $\det B\neq 0$, and
%Q $B_{kl}$
%Q and their derivatives are bounded functions when $y\to x_{0}$,
%Q
we  find
$A(x_{0})=B, \ \partial A/\partial x^{k}{\mid }_{x_{o}} = - \Gamma
_{k}$B.
Hence $\Gamma _{k}A(x_{0}) + \partial A/\partial x^{k}{\mid
}_{x_{o}}\equiv
0$ for  all  $A$ defined  above, i.e.  the  set of vector fields
$\{E^{\prime }_{m} =
A^{i}_{m}\partial /\partial x^{i}\}$ with $[ A^{i}_{m}] = A$
have the needed property. By lemma~\ref{Lem4.1} there exists a
neighborhood
$U$ of $x_0$ such that $\{E_{m}^{\prime}\left.\right|_y$ is a basis
at every
$y\in U$. Hence $\{E_{m}^{\prime}$ form a field of normal bases in
$U$.

\par
Conversely, let $W^{\prime }_{X}(x_0)=0$ in some $\{E^{\prime }_{m}\}$
and every X.
%Q Fixing a basis
%Q $\{E_{i}=(A^{-1})^{m}_{i}E^{\prime }_{m}\}$,
%Q from (\ref{5prime}) we get $W_{X}(x_{0})A(x_{0})+X(A){\mid
%Q }_{x_{0}}=0$,
%Q i.e.
At  $x_0$ from~(\ref{5prime}) we get
$W_{X}(x_{0})=-X(A){\mid }_{x_{0}}A^{-1}(x_{0})=
\Gamma_{k}X^{k}(x_{0})$ for $\Gamma _{k} =
- E_{k}(A){\mid }_{x_{0}}A^{-1}(x_{0})$.~\qed

\par
From (\ref{5prime}) we see that the  normal frames at a given point
are
obtained from one another by linear transformations whose matrices
are such
that  the action of the basic vectors  on them vanish at the given
point.

\par
It follows from the definition of the torsion (see Sec. \ref{II})
that if for an $S$-de\-rivation  there  is  a  local  holonomic
normal frame at $x_{0}$, then its torsion
is zero at $x_{0}$. Conversely, if the torsion vanishes at $x_{0}$ and
normal frames exist, then all of them are holonomic at $x_{0}$.

\par
Due to~(\ref{6}) the linear connections are derivations which are
linear at
every point at which they are defined. Hence, by
proposition~\ref{p4}, for
any linear connection at any point there are normal frames in which
$\Gamma_{jk}^{i}X^k=0$
for every vector field $X$, so in a normal frame $\Gamma_{jk}^{i}=0$.
Thus, the components of a linear connection in a normal frame at a
given
point vanish at that point. For symmetric linear connections this is
a known
result~\cite{3,4,6}, but for nonsymmetric ones this is a new one
established
in 1992 in~\cite{B-1} and reestablished independently in 1995
in~\cite{Hartley}.

\section {VALIDITY OF THE EQUIVALENCE \protect\newline PRINCIPLE}
\label{V}

\par
The above results are physically important in connection with the
equivalence principle. According  to  it (see~\cite{1,2,11} and the
references therein), at  least  at  a point, the laws of special and
general
relativity coincide in a suitably chosen frame.

\par
Let us consider a gravitational theory in which locally the
gravitational
field strength is identified with the local components of
some ($S$-)derivation. The equivalence principle for such a theory,
when  applied  on some set $U\subset M$, demands the field strength to
be (locally) transformable to zero on $U$. Mathematically this means
the
(local) existence of a field of basis (or bases) on $U$ in which  the
components of the mentioned ($S$-)derivation vanish on $U$. As this
work deals
with the cases when $U$ is a single point or a neighborhood  of  a
point, the
following conclusions can be made:

\par
(1) All gravitational theories based on space-times endowed with  a
linear connection (e.g. the general  relativity  \cite{1}  and  the
$U_{4}$
theory \cite{2}) are compatible with the equivalence principle  at
any
fixed space-time point. So, at  any  point  there  exist  (local)
inertial frames, which are holonomic iff the connection is torsion
free
(as  is,  e.g.,  the case of general relativity \cite{1}).

\par
(2) Any gravitational theory based on  space-time  endowed  with  a
linear connection is compatible with the equivalence principle  in
a neighborhood iff  the  connection  is
flat in it. In particular, for flat linear  connections
for \hbox{every} point there exist neighborhoods in which there exist
(local) inertial frames (bases) that are holonomic iff the connection
is torsion free.

\par
(3) The equivalence principle is important when one tries  to
formulate gravi\-tational theories on the base of some (class of)
derivations.  Generally, this principle will select the theories
based on
linear connections (cf.~\cite{11}).

\par
(4) In the above cases the minimal-coupling principle~\cite{1,2},
mathematically realizing the  equivalence  principle  in  any
gravitational theory, looks alike in those  gravitational  theories.
For
instance, if there is also a metric, it  can  be  carried  out  as
outlined in \cite{2}. A physical  law  obtained by means of the
minimal coupling
principle in the considered cases identically  satisfies the
equivalence
principle  as  a  consequence  of  the  underlying mathematics of the
corresponding gravitational theories.

\par
These conclusions, when applied to the case of general relativity,
are in
full agreement with the ones of~\cite{Fock}, where it is argued that
the
equivalence principle is a theorem in general relativity. But our
results are
far more general, in particular, they are valid for any gravitational
theory
based on linear connections with or without torsion.

\section {CONCLUSION}
\label{VI}

\par
The linear connections, as we have seen, are remarkable among
all derivations with their property that in a number of sufficiently
general
cases considered here   they  are  the  only  derivations admitting
special
bases in which their components vanish.

\par
This formalism seems applicable also to fields different from
the gravitational one, viz. at least to those described by  linear
connections. This suggests the idea of extending the validity of the
equivalence principle outside the gravitational interaction (cf.
\cite{kapus}).

\par
Elsewhere this formalism will be generalized along paths or on
more general submanifolds of the space-time.

\section * {ACKNOWLEDGEMENTS}

\par
The author is grateful to Professor Stancho Dimiev (Institute
of Mathematics of Bulgarian Academy of Sciences) and to Dr.  Sava
Manoff (Institute for Nuclear Research and Nuclear Energy of
Bulgarian
Academy of Sciences) for the valuable comments and stimulating
discussions.

\end{document}